\documentclass[12pt,twoside,dvips]{article}
\usepackage{amssymb}
\usepackage{amsfonts}
\usepackage{amsmath}
\usepackage{pifont}
\usepackage[spanish,english]{babel}
\usepackage{graphicx}

\begin{document}

\title{A 3-3-1 model with SU(8) unification}
\author{R. Mart\'{\i}nez$^{1}\thanks{%
e-mail: remartinezm@unal.edu.co}$, F. Ochoa$^{1}\thanks{%
e-mail: faochoap@unal.edu.co}$, P. Fonseca$^{1,2}$ \and $^{1}$Departamento de F\'{\i}sica, Universidad
Nacional de Colombia, \\ Ciudad Universitaria, Bogot\'{a} D.C. \and $^2$Departamento de F\'{\i}sica, Universidad Pedag\'{o}gica y Tecnol\'{o}gica de Colombia, \\ Avenida Central del Norte, Tunja}
 
\maketitle

\begin{abstract}
We construct a 3-3-1 model for three families that can be embedded into a single SU(8) unified model. Assuming appropriate branching rules and symmetry-breaking pattern, we find a complete fermion content within irreducible representations of SU(8), where light standard model fermions, heavy 3-3-1 fermions and super-heavy fermions may be distinguished. In the framework of the doubly lopsided mechanism, we obtain mass matrix structures which exhibit nontrivial flavor hierarchical features. Among the up-type quarks, one (top quark) has tree-level mass and two (charm and up) get masses at one-loop level. Considering only dominant contributions, we may obtain the ratios $|m_c/m_t| \approx 7.4 \times 10^{-3}$ and $|m_u/m_c| \approx 1.9 \times 10^{-3}$ with few assumptions on the free parameters and without any hierarchical requirements on the Yukawa couplings.              
 
\end{abstract}

\section{Introduction}

Although the Standard Model (SM) is the simplest model that succesfully explain most of the phenomena and experimental observations in particle physics, it contains unanswered fundamental questions which many theorists associate to an underlying unified theory beyond the SM. For example, in the framework of the SM the enormous differences in the scale of masses exhibited among the elementary fermions is unnatural and unpredictable. This question can be addressed in family dependent models where a symmetry distinguish fermions of different families. An interesting alternative that may provide a clue to this puzzle are the models with gauge symmetry $SU(3)_c \otimes SU(3)_L \otimes U(1)_X$, also called 3-3-1 models, which introduce a family non-universal $U(1)$ symmetry \cite{331-pisano, 331-frampton, 331-long, M-O}. These models have a number of phenomenological advantages. First of all, from the cancellation of chiral anomalies \cite{anomalias} and
asymptotic freedom in QCD, the 3-3-1 models can explain why there are three fermion
families. Secondly, since the third family is treated under a different representation, the
large mass difference between the heaviest quark family and the two lighter ones may be
understood \cite{third-family}. Third, the models have a scalar content similar to the two Higgs doublet
model (2HDM), which allow to predict the quantization of electric charge and the vectorial
character of the electromagnetic interactions \cite{quantum-charge, vectorlike}. Also, these models contain a natural Peccei-Quinn symmetry, necessary to solve the strong-CP problem \cite{PC, PC331}. Finally, the model introduces new types of matter relevant to the next generation of colliders at the
TeV energy scales, which do not spoil the low energy limits at the electroweak scale.

Although the 3-3-1 models introduce a family non-universal property, it is exhibited only for the quark sector between the heaviest family and the remaining two lighter. Thus, the model does not reproduce well the hierarchical strutures of the quark/lepton mass matrices. In addition, none of the current non-supersymetrical 3-3-1 models can be considered as an unification model at the Planck scale but a model at an intermediate scale of energy. An alternative to connect 3-3-1 models with unification theories is to embed them into bigger gauge structures and to generate an appropriate spontaneuos symmetry breaking mechanism. Although there are some attempts to achieve this with 3-3-1 models, there arise problems that either spoils some of the good features of the model or it simply does not improve the predictive power of the model. Thus, for example, some authors have considered 3-3-1 models as a subgroup of an $SU(6) \otimes U(1)$ model \cite{SU6} and $E6$ groups  \cite{E6}, which is possible only for the unrealistic case of one family. Other authors have proposed to embed 3-3-1 models into the Pati-Salam $SU(4)_ps \otimes SU(4)_{L+R}$ model \cite{4-4}, which exhibit mass matrices analogous to the 3-3-1 case and without the unification of the interactions. There are other extensions inspired in the 3-3-1 models as for example $SU(3)_c \otimes SU(4)_L \otimes U(1)$ models \cite{341}, which basically exhibit the same features as the 3-3-1 case, without neither explaining the hierarchical pattern of the mass structures nor to achive an unification of the interactions.

On the other hand, there is another mechanism to generate hierarchical pattern of the fermion masses without a flavor symmetry, based in the old idea that in a three-family grand unified gauge model, each family transform differently under the unified group, producing a nontrivial flavor structure \cite{GUT-hier}. The combination of this scheme with that where the masses of light fermions may arise out of radiative corrections whereas the heavy fermions obtain tree-level masses \cite{rad-mass}, leads to realistic texture structures for the quark and lepton mass matrices. Recently, these schemes have been reviewed and new ideas have been introduced by authors in \cite{rev-GUT, barr} in the context of SO(10)- and SU(N)-based models, and in \cite{framp} for SU(9) models. For example, in SU(N) models the way mass hierarchy emerges is characteristic of doubly lopsided models originally proposed in the framework of atmospheric neutrinos \cite{lopside}. Although the above works show that it is possible to obtain predictive flavor structures without flavor symmetry in a grand unification scheme, and applications to specific models have been described, the constructions and analyses have been done in a qualitative level, where general assumptions on the particle content and on the scheme of the breaking symmetry are considered. We intend to explore in a quantitative level these methods within a unified model with a specific particle content and breaking symmetry sequence. In contrast with the previous works where the SU(5) "language" is used to descompose the irreducible representations of SU(8), we use a 3-3-1 language to identify particles within SU(8) representations. In this paper we emphasize in the theoretical construction of the model, where we describe how to embed 3-3-1 into SU(8) and a complete particle assignation into irreducible representations (irreps) of SU(8) is obtained. We show that it is possible to construct a 3-3-1 model for three families that can be unified into a single unified group, and symmetry-breaking pattern from SU(8) down to 3-3-1 may be generated. Detailed numerical and phenomenological analyses including studies on the renormalization group equations, construction of Higgs potential, one-loop calculations, etc. is intended to be done in future works. However, in order to illustrate the way mass hierarchy emerges from this model, one-loop diagrams for the up-quark sector are constructed and taking only dominant constributions, we estimate the hierarchical ratios among their masses. In sec. 2 we show all the group structure and representations of SU(8), and from appropriate branching rules we generate decompositions of representations down to 3-3-1. We obtain a complete three-family fermion 3-3-1 spectrum. In Sec. 3 we exhibit the complete SU(8) particle content of the model. Secs. 4 and 5 are devoted to the discussion of the fermion mass generation. In 4 we show the most general Yukawa terms that induce tree-level matrix masses, while in 5 we obtain an approximation of one-loop mass calculations for the up-type quark sector.               
  

\section{Embedding 331 into SU(8)}

We requiere that the unified group SU(8) contains an $SU(3)_c \otimes G_{fl}$ with $c$ the color charge of the Quantum Chromodynamics (QCD) and $fl$ the flavor sector of the electroweak interactions to be identified with the $SU(3)_L \otimes U(1)_X$ group. Let us consider the following breakdown chain of SU(8) to a 3-3-1 group:

\begin{align}
\nonumber SU(8) &\rightarrow SU(4)_c \otimes SU(4)_L \otimes U(1)_I \\ \nonumber
& \rightarrow SU(3)_c  \otimes SU(4)_L \otimes U(1)_c \otimes U(1)_I \\ 
&\rightarrow SU(3)_c  \otimes SU(3)_L \otimes U(1)_c \otimes U(1)_L \otimes U(1)_I. 
\label{embed}
\end{align}

In order to identify family structures and sub-group generators, we use in Eq. (\ref{embed}) the following branching rules for the fundamental irreducible representation (irrep) of SU(8): 
 
\begin{align}
\nonumber 8 &\rightarrow (4,1)(-1)+(1,4)(1) \\ \nonumber
&\rightarrow (3,1)(1/3,-1)+(1,1)(-1,-1)+(1,4)(0,1) \\ 
&\rightarrow (3,1)(1/3,0,-1)+(1,1)(-1,0,-1)+(1,3)(0,1/3,1)+(1,1)(0,-1,1),
\label{bran-rule}
\end{align}

where we use the notation (x,y)(a,b,..), with (x,y) the irreps of $(SU(n),SU(m))$ subgroups in the same order as shown in (\ref{embed}), and the second parenthesis (a,b,..) contains the unnormalized U(1) charges. From the branching rules in (\ref{bran-rule}), we can identify the following U(1) generators:

\begin{align}
\nonumber U(1)_c &\rightarrow X_c = N_c diag(1/3,1/3,1/3,-1,0,0,0,0), \\ \nonumber
U(1)_L &\rightarrow X_L = N_L diag(0,0,0,0,1/3,1/3,1/3,-1), \\ 
U(1)_I &\rightarrow X_I = N_I diag(-1,-1,-1,-1,1,1,1,1),
\label{U1-gener}
\end{align}  

where the components $(1,2,3)$ and $(5,6,7)$ are associated to $SU(3)_c$ and $SU(3)_L$ triplets, respectively, while the components $(4,8)$ are singlets. $N_I, N_L$ and $N_c$ are appropriate normalization factors. For the last breakdown in (\ref{embed}) we put together the U(1)'s charges into a unique $U(1)_X$ group, where:

\begin{equation}
X=\alpha _1 (1/N_I )X_I + \alpha _2 (1/N_c )X_c + \alpha _3 (1/N_L )X_L.
\label{X-charge}
\end{equation}

The resulting $SU(3)_c \otimes SU(3)_L \otimes U(1)_X$ subgroup is broken down further to the SM gauge group:

\begin{align}
\nonumber SU(3)_c \otimes SU(3)_L \otimes U(1)_X &\rightarrow SU(3)_c \otimes SU(2)_L \otimes U(1)_Y  \\
 &\rightarrow (3,1)(Y_c )+(1,1)(Y_4 )+(1,2)(Y_L)+(1,1)(Y_7 )+(1,1)(Y_8),
\label{331-break}
\end{align}

where $Y_c$ and $Y_L$ are the weak hipercharges of the components $(1,2,3)$ and $(5,6)$, respectively, while $Y_{4,7,8}$ are the hipercharges of the $4^{th}$, $7^{th}$ and $8^{th}$ components. These hipercharges and the electric charges $Q$ are related by

\begin{equation}
Q=T_3 + Y= T_3 + \beta T_8 + X,
\label{elec-charge}
\end{equation}

with $T_3 =$ $(1/2)Dg(0,0,0,0,1,-1,0,0)$ and $T_8 =($1/2$\sqrt{3})$$Dg(0,0,0,0,1,1,-2,0)$ the diagonal $SU(3)_L$ generators embedded into SU(8). $\beta$ is a free parameter to be determined by the particle content of the model. In order to avoid exotic charges, at least in the 3-3-1 sub-sector that will arise, we choose $\beta = 1/\sqrt{3}$ \cite{331-long, M-O}. From Eq. (\ref{331-break}) we see a (3,1) representation under $(SU(3)_c , SU(2)_L)$ which can be identified with a quark weak singlet, and a (1,2) representation associated with lepton weak doublets. For instance, we can indentify in (\ref{331-break}) the SM case with $Y_c = -1/3$ and $Y_L = 1/2$, corresponding to a right-handed down quark in (3,1), and a right-handed $(e^{+}, \bar{\nu _{e} })$ doublet for (1,2), respectively. With the above choose applied into Eqs. (\ref{U1-gener})-(\ref{elec-charge}), we obtain for the $\alpha$ coefficients from (\ref{X-charge}) that:

\begin{align}
3\alpha _1-\alpha _2=1,  \hspace{1cm} 3\alpha _1+\alpha _3=1.
\label{alphas}
\end{align}

The above relations fix $Y_7 = 0$ for the second singlet (1,1) in Eq. (\ref{331-break}). For the other two (1,1) singlets, we also require that there are not exotic electric charges, i.e. $Q_{4,8}= \pm{1}$ or $0$. These choose combined with Eqs (\ref{alphas}) and (\ref{U1-gener})-(\ref{elec-charge}) lead us to three posibilities:

\begin{align}
\nonumber &(\mathcal{A}) \hspace{0.5cm} \alpha _1 =-\alpha _2 = \alpha _3 =1/4, \\ \nonumber
&(\mathcal{B}) \hspace{0.5cm} \alpha _1 =\alpha _2 =-\alpha _3 =1/2, \\
&(\mathcal{C}) \hspace{0.5cm} \alpha _1 = 0, \alpha _2 =-\alpha _3 = -1.
\label{alfas}
\end{align}

On the other hand, the fundamental irreducible representation of SU(8) is not enough to fit all of the SM fermion content. Thus, it is necessary to include other representations. In particular, in order to avoid $SU(3)_c$ irreps bigger than triplets and anti-triplets, we should restrict our options to the totally antisymmetric irreps of rank $[p]$. In $SU(8)$ we find 7 antisymmetric irreps corresponding to tensors with rank $[1],[2],[3],[4],\overline{[3]},\overline{[2]}$ and $\overline{[1]}$, with dimensions $\mathbf{8, 28, 56, 70, 56^{\ast}, 28^{\ast}, 8^{\ast}}$, respectively, where $\overline{[p]}$ denotes conjugate tensors. As noted by authors in ref. \cite{barr}, the most economical and anomaly free set of irreps is $9\overline{[1]}+[2]+[3]$, i.e. nine conjugate tensors of rank 1 ($\psi _{(m)A}$), one tensor of rank 2 ($\psi ^{[AB]}$) and one of rank 3 ($\psi ^{[ABC]}$), where $m=1,2,..,9$ is the index that label the nine tensors of rank 1, and $A,B,C=1,..,8$ are the SU(8) indices. Taking into account that tensors with rank bigger than one transform as direct product of fundamental representations, we find the branching rules of the [2] and [3] irreps. For the breakdown from (\ref{embed}), we find the following decompositions under $SU(3)_c \otimes SU(3)_L$:

\begin{align}
\nonumber 9 \times \overline{[1]}=\psi _{(m)A} &\rightarrow \psi _{(m)a}+ \psi _{(m)4}+ \psi _{(m)i}+\psi _{(m)8} \\
\nonumber 9 \times 8^* &\rightarrow 9 \times [(3^* ,1)+(1,1)+(1,3^* )+(1,1)] \\
\nonumber [2]= \psi ^{[AB]} &\rightarrow \psi ^{[ab]} + \psi ^{[a4]} + \psi ^{[ai]} + \psi ^{[a8]} + \psi ^{[4i]} + \psi ^{[48]} + \psi ^{[ij]} + \psi ^{[i8]} \\
\nonumber 28 &\rightarrow (3^* ,1)+(3,1)+(3,3)+(3,1)+(1,3)+(1,1)+(1,3^* )+(1,3) \\
\nonumber [3]= \psi ^{[ABC]} &\rightarrow \psi ^{[abc]} + \psi ^{[ab4]} + \psi ^{[abi]} + \psi ^{[ab8]} + \psi ^{[a4i]} + \psi ^{[a48]} +\psi ^{[aij]} + \psi ^{[ai8]} +\psi ^{[4ij]} \\ \nonumber & + \psi ^{[4i8]} + \psi ^{[ijk]} + \psi ^{[ij8]} \\
\nonumber 56 &\rightarrow (1,1)+(3^* ,1)+(3^* ,3)+(3^* ,1)+(3,3)+(3,1)+(3,3^*)+(3,3)+(1,3^* ) \\ &+(1,3) +(1,1)+(1,3^* ),
\label{irrep-bran}
\end{align}

where $a,b,c,..=(1,2,3)$ label $SU(3)_c$ indices, $i,j,k,..=(5,6,7)$ label $SU(3)_L$ indices and $4,8$ are singlets. In order to identify a low energy 3-3-1 spectrum into the representations in Eq. (\ref{irrep-bran}) (i.e, those representations that still remains massless after the breakdown from Eq. (\ref{embed})), we choose the minimal anomaly free set of triplets, antitriplets and singlets that contains as subset all the SM particles. In order to make the particle assignation, we first obtain the electric charges of the representations in (\ref{irrep-bran}) using the definitions from eqs. (\ref{U1-gener}),(\ref{X-charge}) and (\ref{elec-charge}), and later we match representations and charges in a convenient form in order to obtain the SM spectrum (plus the new 3-3-1 particles). We find the charges shown in Eq. (\ref{Ap:charges1})-(\ref{Ap:charges3}) of the Appendix \ref{apendixA}. For the $\alpha$ coefficients, we are restricted to the three Eqs. in (\ref{alfas}). We see that the second option leads to the exotic charge $Q^{[a4]}=-4/3$ in Eq. (\ref{Ap:charges2}), which we intend to avoid. Thus, we discard solution $(\mathcal{B})$ in (\ref{alfas}). We also see that solution $(\mathcal{C})$ leads to an exotic quark with charge $Q^{[ab8]}=-5/3$. Then, we obtain that $(\mathcal{A})$ is the only solution without exotic electric charges, obtaining the following charge structures:

\begin{align}
\nonumber \psi _{(m)A} \rightarrow Q_{[a]}&=1/3, \hspace{0.2cm} Q_{[4]}=0, \hspace{0.2cm} Q_{[i]}=\left(-1,0,0 \right), \hspace{0.2cm} Q_{[8]}= 0 \\  \nonumber
\psi ^{[AB]} \rightarrow Q^{[ab]}&=-2/3, \hspace{0.2cm} Q^{[a4]}=-1/3, \hspace{0.2cm} Q^{[ai]}=\left(2/3,-1/3,-1/3  \right), \hspace{0.2cm} Q^{[a8]}= -1/3 \\ \nonumber
 Q^{[4i]}&=\left(1, 0, 0 \right), \hspace{0.2cm} Q^{[48]}=0 , \hspace{0.2cm} Q^{[ij]}=\left(1, 1, 0 \right) \hspace{0.2cm} Q^{[i8]}=\left(1, 0, 0 \right) \\ \nonumber 
\psi ^{[ABC]} \rightarrow Q^{[abc]}&=-1, \hspace{0.2cm} Q^{[ab4]}=-2/3, \hspace{0.2cm} Q^{[abi]}=\left(1/3, -2/3, -2/3 \right), \hspace{0.2cm} Q^{[ab8]}=-2/3, \\ \nonumber
 Q^{[a4i]}&=\left(2/3, -1/3, -1/3 \right), \hspace{0.2cm} Q^{[a48]}=-1/3 , \hspace{0.2cm} Q^{[aij]}= \left(2/3, 2/3, -1/3 \right), \\ \nonumber
 Q^{[ai8]}&=\left(2/3, -1/3, -1/3 \right), \hspace{0.2cm} Q^{[4ij]}=\left(1, 1, 0 \right), \hspace{0.2cm} Q^{[4i8]}=\left(1, 0, 0 \right), \\
 Q^{[ijk]}&=1, \hspace{0.2cm} Q^{[ij8]}=\left(1, 1, 0 \right) 
\label{charges}
\end{align}

where the charges in triplet arrangements are associated with the flavor indices $i,j,k=5,..7$. Taking left-handed fermions, we should fit at least the following SM three-family structures: three anti-down quarks $d^{c}_{(n)L}$ (we use the notation $3\times (1/3)$ for three charges of 1/3 and $n=1,2,3$ label each family), $3\times(-1/3)$ for the corresponding down-type quark families $d_{(n)L}$, $3\times (-1)$ for electron-type flavors $e^{-}_{(n)L}$, $3\times (+1)$ for the corresponding positrons $e^{+}_{(n)L}$, $3\times (0)$ for 3 light majorana neutrinos $\nu _{(n)L}$, $3\times (-2/3)$ for anti-up families $u^{c}_{(n)L}$, and $3\times (2/3)$ for the corresponding up-type quarks $u_{(n)L}$. Thus, we should match these charges with the charges in Eq. (\ref{charges}) and representations in (\ref{irrep-bran}). We proceed in the following way:

\vspace{0.3cm}

$\mathit{1}-$ The anti down-type quark sector are assigned in three of the nine $\psi_{(m)a}$. We choose $m=1,2,3=n$ for $d^{c}_{(n)L}$ (note that it match the correct charge of 1/3 with $Q_{[a]}$ from (\ref{charges})).

\vspace{0.2cm}

$\mathit{2}-$ The electron-type leptons and neutrinos are assigned into $\psi_{(m)i}$. However, these embed into anti-triplet arrangements with charges $Q_{[i]}=(-1,0,0)$, obtaining an extra majorana neutral lepton labeled as $N^{0}_{(n)L}$. Choosing again $m=1,2,3=n$ for three families, we get $\psi_{(n)i} \rightarrow (e^{-}_{(n)},\nu _{(n)}, N^{0}_{(n)})_L$. 

\vspace{0.2cm}

$\mathit{3}-$ In (\ref{charges}) we find only three $-2/3$ singlet charges associated to the representations $\psi ^{[ab]},\psi ^{[ab4]}$ and $\psi ^{[ab8]}$. These are good candidates to match with anti-up singlet quarks $u^{c}_{(n)L}$. 

\vspace{0.2cm}

$\mathit{4}-$ In (\ref{irrep-bran}) we find only three (3,3) representations associated with $\psi ^{[ai]},\psi ^{[a4i]}$ and $\psi ^{[ai8]}$, which contain charges $(2/3,-1/3,-1/3)$ each. Thus, we can fit the three families of up- and down-type quarks, plus new down-type quarks, obtaining the triplets $(u_{(n)}, d_{(n)}, D_{(n)})_L$. The new $D_{(n)L}$ fields need corresponding conjugate fields with charges of $1/3$ in order to construct Dirac fermions. As we did with  $d^{c}_{(n)L}$, we indentify $D^{c}_{(n)L}$ with three of the nine $\psi_{(m)a}$ tensors. We choose $m=4,5,6$ for these fields.   

\vspace{0.2cm}

$\mathit{5}-$ We could match the positrons with flavor singlets. However, we see that there are not enough $(1,1)$ representations with the correct charges in order to fit the three positron families. On the other hand, we find three $(1,3^{*})$ representations corresponding to the $\psi ^{[ij]},\psi ^{[4ij]}$ and $\psi ^{[ij8]}$ tensors with charges $(1,1,0)$ each. Thus, if we include new charged and neutral leptons, we can construct the triplets $(e^{+}_{(n)}, E^{+}_{(n)}, \tilde{E} ^{0} _{(n)})_L$. The new charged leptons $E^{+}_{(n)}$ need conjugate leptons in order to form Dirac fermions. As done with the electrons $e^{-}_{(n)}$, we identify $E^{-}_{(n)L}$ with three of the $\psi_{(m)i}$ tensors. We choose $m=7,8,9$. However, $\psi_{(m)i}$ also form triplets with charges $(-1,0,0)$ as seen in ($\ref{charges}$), which introduce two other neutral majorana leptons that we call $E^{0}_{(n)}$ and $F^{0}_{(n)}$.

\vspace{0.3cm}

In conclusion, we obtain the following three-family low energy 3-3-1 spectrum embedded into SU(8):

\begin{align}
\nonumber \ell _{(n)L}&=\left(
\begin{array}{c}
e^{-}_{(n)} \\ 
-\nu _{(n)} \\ 
N^{0}_{(n)}
\end{array}
\right) _{L}:\left(\mathbf{1},\mathbf{3}^{\ast }\right)\left(-1/3\right), \\ \nonumber
\mathcal{N} _{(n)L}&=\left(
\begin{array}{c}
E^{-}_{(n)} \\ 
-E^{0} _{(n)} \\ 
F^{0}_{(n)}
\end{array}
\right) _{L}:\left(\mathbf{1},\mathbf{3}^{\ast }\right)\left(-1/3\right), \\
\mathcal{M} _{(n)L}&=\left(
\begin{array}{c}
E^{+}_{(n)} \\ 
-\tilde{E} ^{0} _{(n)} \\ 
e^{+}_{(n)}
\end{array}
\right) _{L}:\left(\mathbf{1},\mathbf{3}^{\ast }\right)\left(2/3\right), 
\label{331-lep}
\end{align}

for leptons, and

\begin{align}
\nonumber Q_{(n)L}&=\left(
\begin{array}{c}
u_{(n)} \\ 
d_{(n)} \\ 
D_{(n)}
\end{array}
\right) _{L}:\left(\mathbf{3},\mathbf{3}\right)\left(0\right), \\ \nonumber
u^{c}_{(n)L}&:\left(\mathbf{3}^{\ast},\mathbf{1}\right)\left(-2/3\right), \\ \nonumber
d^{c}_{(n)L}&:\left(\mathbf{3}^{\ast},\mathbf{1}\right)\left(1/3\right), \\ 
D^{c}_{(n)L}&:\left(\mathbf{3}^{\ast},\mathbf{1}\right)\left(1/3\right),
\label{331-quark}
\end{align}

for quarks. The fermions transform under $(SU(3)_c, SU(3)_L)(U(1)_X)$ as shown in the above equations, where the $X$ charges were calculated through Eq. (\ref{X-charge}). Thus, we find all the SM particles plus new 3-3-1 particles for three families embedded into a SU(8) group. Although the spectrum from (\ref{331-lep}) and (\ref{331-quark}) exhibit a family universal structure in the $3-3-1$ subgroup, they come from different representations in $SU(8)$, thus leading to a family hierarchy, as we will see below.  

\section{The complete SU(8) spectrum}

From the charge values found in (\ref{charges}) and taking into account the 3-3-1 spectrum found above, we write all the fermion spectrum contained into the irreps $\psi _{(m)A}$, $\psi ^{[AB]}$ and $\psi ^{[ABC]}$ of SU(8). After the analysis of all quantum numbers, we identify for the irrep $\mathbf{8}^{\ast}$ the following structures:

\begin{align}
 \psi _{(m)A} \rightarrow \psi _{(p)a} &= d^{c}_{(n)aL},& \hspace{0cm} \psi _{(q)a} &= D^{c}_{(n)aL},&\hspace{0cm} \psi _{(r)a} &= J^{c}_{(n)aL}& \nonumber \\ 
\psi _{(p)4} &= X^{0}_{(n)L},& \hspace{0cm} \psi _{(q)4} &= R^{0}_{(n)L},& \hspace{0cm} \psi _{(r)4} &= B^{0}_{(n)L}& \nonumber \\ 
\psi _{(p)i} &= \left(e^{-}_{(n)},\nu _{(n)},N^{0}_{(n)}\right)_L, & \hspace{0.1cm} \psi _{(q)i} &= \left(E^{-}_{(n)},E^{0}_{(n)},F^{0}_{(n)}\right)_L,& \hspace{0.1cm} \psi _{(r)i} &= \left(C^{-}_{(n)},C^{0}_{(n)},G^{0}_{(n)}\right)_L & \nonumber \\ 
\psi _{(p)8} &= Y^{0}_{(n)L},& \hspace{0cm} \psi _{(q)8} &= P^{0}_{(n)L},& \hspace{0cm} \psi _{(r)8} &= I^{0}_{(n)L},& 
\label{8-spectrum}
\end{align}  

where $m=1,..,9$ splits in three "horizontal" structures label with $p=1,2,3$, $q=4,5,6$ and $r=7,8,9$, while $n=1,2,3$ label families into each structure. For the structure $\psi _{(m)a}$, the "vertical" index $a=1,2,3$ run over color charges, while for $\psi _{(m)i}$, the index $i=5,6,7$ run over flavor indices which are written explicitly as triplets. The $\psi _{(m)4,8}$ components are color and flavor singlets.

For the irrep $\mathbf{28}$, we get: 

\begin{align}
\nonumber \psi ^{[AB]} \rightarrow \psi ^{[ab]} &=u^{c}_{(1)aL} \\ \nonumber
\psi ^{[a4]} &=J_{(1)aL} \\ \nonumber
\psi ^{[ai]} &=\left(u_{(1)a}, d_{(1)a}, D_{(1)a}\right)_{L} \\ \nonumber
\psi ^{[a8]} &=J_{(2)aL} \\ \nonumber
\psi ^{[4i]} &=\left(C^{+}_{(1)}, L^{0}_{(1)}, M^{0}_{(1)}\right)_{L} \\ \nonumber
\psi ^{[48]} &=O^{0}_{L} \\ \nonumber
\psi ^{[ij]} &=\left(e^{+}_{(1)}, E^{+}_{(1)}, \tilde{E}^{0}_{(1)}\right)_{L} \\
\psi ^{[i8]} &=\left(C^{+}_{(2)}, L^{0}_{(2)}, M^{0}_{(2)}\right)_{L} 
\label{28-spectrum}
\end{align}

where $(1)$ and $(2)$ label families 1 and 2, respectively.

For the irrep $\mathbf{56}$ we have:

\begin{align}
\nonumber \psi ^{[ABC]} \rightarrow \psi ^{[abc]}&=S^{-}_{L} \\ \nonumber
\psi ^{[ab4]}&=u^{c}_{(2)aL} \\ \nonumber
\psi ^{[abi]}&=\left(T^{c}_{a},K^{c}_{a},V^{c}_{a}\right)_{L} \\ \nonumber 
\psi ^{[ab8]}&=u^{c}_{(3)aL} \\ \nonumber
\psi ^{[a4i]}&=\left(u_{(2)a},d_{(2)a},D_{(2)a}\right)_{L} \\ \nonumber
\psi ^{[a48]}&=J_{(3)aL} \\ \nonumber
\psi ^{[aij]}&=\left(V_{a},K_{a},T_{a}\right)_{L} \\ \nonumber
\psi ^{[ai8]}&=\left(u_{(3)a},d_{(3)a},D_{(3)a}\right)_{L} \\ \nonumber
\psi ^{[4ij]}&=\left(e^{+}_{(2)},E^{+}_{(2)},\tilde{E}^{0}_{(2)}\right)_{L} \\ \nonumber
\psi ^{[4i8]}&=\left(C^{+}_{(3)},L^{0}_{(3)},M^{0}_{(3)}\right)_{L} \\ \nonumber
\psi ^{[ijk]}&=S^{+}_{L} \\
\psi ^{[ij8]}&=\left(e^{+}_{(3)},E^{+}_{(3)},\tilde{E}^{0}_{(3)}\right)_{L}
\label{56-spectrum}
\end{align}

where $(3)$ label the third family. The corresponding electric charges of all the above spectrum can be read from Eq. (\ref{charges}). We can classify the above spectrum in light fermions (the ordinary SM particles), heavy fermions (the new 3-3-1 particles writen in (\ref{331-lep}) and (\ref{331-quark})), and superheavy fermions (any other particle different from ordinary and 3-3-1 fermions). In particular, we can identify the 3-3-1 particle content from (\ref{331-lep}) and (\ref{331-quark}) embedded into the above spectrum but with families localized in different SU(8) representations, which produces a nontrivial flavor structure. 

\section{Yukawa Lagrangian}

With the irreps in Eq. (\ref{irrep-bran}), the only allowed renormalizable Yukawa terms in SU(8) have the following tensor structures:

\begin{align}
\nonumber \overline{[1]}_{f} \overline{[1]}_f[2]_H= &a_{ml}\overline{\left(\psi _{(m)A}\right) ^{c}}\psi _{(l)B}H^{[AB]}, \\ \nonumber
\overline{[1]}_f[2]_f\overline{[1]}_H=&Y_{m} \overline{\left(\psi _{(m)A}\right) ^{c}}\psi ^{[AB]}H^{\ast}_{B} , \\ \nonumber
\overline{[1]}_f[3]_f \overline{[2]}_H=&y_{m} \overline{\left(\psi _{(m)A}\right) ^{c}}\psi ^{[ABC]}H^{\ast}_{[BC]} , \\ \nonumber
[2]_f[2]_f \overline{[4]}_H=&Y \overline{\left(\psi ^{[AB]}\right) ^{c}}\psi ^{[CD]}H^{\ast}_{[ABCD]} , \\ 
[2]_f[3]_f \overline{[5]}_H=&y \overline{\left(\psi ^{[AB]}\right) ^{c}}\psi ^{[CDE]}H^{\ast}_{[ABCDE]} ,
\label{yukawa} 
\end{align} 

where we have introduced appropriate Higgs fields in antisymmetric representations $H^{\ast}_{A}=\mathbf{8}^{\ast}$, $H^{[AB]}=\mathbf{28}$, $H^{\ast}_{[ABCD]}=\mathbf{70^{\ast}}$ and $H^{\ast}_{[ABCDE]}=\varepsilon _{ABCDEFGH}H^{FGH}=56^{\ast}$. The Yukawa constant $a_{ml}$ is antisymmetric. The tensor structure of the form $[3][3][2]$ is canceled out by the product of antisymmetrical tensors. The Vacuum Expectation Values (VEV) of the Higgs fields will depend on two conditions: i.) they should break the appropriate generators in order to induce the symmetry breaking from Eqs. (\ref{embed}) and (\ref{331-break}), and ii.) the mass terms that will arise from (\ref{yukawa}) should respect the $U(1)_Q$ electric charge conservation after the particle assignation from (\ref{8-spectrum})-(\ref{56-spectrum}) have been done. For the first condition, we introduce the following hierarchical breakdown scales:

\begin{align}
\nonumber SU(8) &\rightarrow SU(4)_c \otimes SU(4)_L \otimes U(1)_I \hspace{0.6cm}(\text{at} \hspace{0.2cm} V_I)\\ \nonumber
& \rightarrow SU(3)_c  \otimes SU(4)_L \otimes U(1)_c \otimes U(1)_I \hspace{0.6cm}(\text{at} \hspace{0.2cm} V_c)\\ \nonumber
&\rightarrow SU(3)_c \otimes SU(3)_L \otimes U(1)_X \hspace{0.6cm}(\text{at} \hspace{0.2cm} V_{jX})\\ \nonumber
&\rightarrow SU(3)_c \otimes SU(2)_L \otimes U(1)_Y \hspace{0.6cm}(\text{at} \hspace{0.2cm} V_{jY})\\
&\rightarrow SU(3)_c \otimes U(1)_Q \hspace{0.6cm}(\text{at} \hspace{0.2cm} \nu _{jw}). 
\label{scales}
\end{align}

where each scale is identified with the following VEVs of the Higgs fields:  

\begin{align}
\nonumber \langle H^{A} \rangle \rightarrow &\langle H^{4}\rangle =V _{c}, \langle H^{6}\rangle =\nu _{1w},\langle H^{7}\rangle =V _{1Y},\langle H^{8}\rangle =V _{1X}, \\ \nonumber
\langle H^{[AB]} \rangle \rightarrow &\langle H^{[46]} \rangle=\nu _{2w}, \langle H^{[47]} \rangle=V_{2Y}, \langle H^{[48]} \rangle=V_{2X}, \langle H^{[67]} \rangle=\nu _{3w},\\ \nonumber &\langle H^{[68]} \rangle=\nu _{4w},\langle H^{[78]} \rangle=V_{3Y}, \\ \nonumber
\langle H^{[ABCD]} \rangle \rightarrow &\langle H^{[1235]} \rangle=\nu _{5w},\langle H^{[4678]} \rangle=\nu _{6w}, \\
\langle H^{[ABCDE]} \rangle \rightarrow &\langle H^{[12345]} \rangle=\nu _{7w}, \langle H^{[12356]} \rangle=V_{4Y}, \langle H^{[12357]} \rangle=\nu _{8w}, \langle H^{[12358]} \rangle=\nu _{9w}.
\label{VEV}
\end{align}

The notation $V_{jX}$, $V_{jY}$ and $\nu _{jw}$ mean the $j$-th VEV term in each of the last three breaking in (\ref{scales}), which arise from the analysis of the symmetry breaking of the group generators. Since the $SU(8)$ generators can not be broken through the Higgs structures chosen in Eq. (\ref{yukawa}), we see that the first breakdown at the scale $V_I$ is not included in the VEVs in (\ref{VEV}). Therefore, it is necessary to include an additional Higgs field in adjoint representation $Adj_H$ with a superheavy diagonal VEV $\langle \Omega \rangle = V_I$. This field can not induce particle masses at tree level, but it can generate effective operators in one-loop diagrams which provide additional mass terms. Although we do not know a priori how the mass scales are constrained, for the sake of the present analysis we suppose that $V_I \gg$ $V_c \sim$ $V_{jX} \gg$ $V_{jY} \gg$ $\nu _{jw}$ to explore how hierarchical mass structures can arise. Since we do not know which of the two SU(4) groups break first at $V_c$ and $V_{jX}$ in (\ref{scales}), we consider that $V_c \sim V_{jX}$. A detailed study on the energy scales of the breakdown of $SU(8)$ require a complete analysis of the renormalization group, which we intend to perform in future works. With the above contraints in the VEVs and with a single $[1]_H$, $[2]_H$, $[4]_H$ and $[5]_H$ set of Higgs fields, we obtain the mass structures at tree level shown in the App. (\ref{appendixAA}), where $\nu _{w}=\nu _{jw}$ for all $j=1,..9$, $V_{Y}=V_{jY}$ for all $j=1,..4$, and $V_c=V_X=V_{jX}$ for $j=1,2$. In the charged lepton sector the mass matrix is given by Eq. (\ref{Ap:char-lep}) which is written in terms of $3 \times 3$ blocks exhibited in Eqs. (\ref{Ap:light-char})-(\ref{Ap:H3-char}). In the block basis $\left(\underline{e},\underline{E},\underline{C}\right)$, we can write down the charged lepton matrix as:

\begin{align}
M^{0}_{l}&=\left(
\begin{array}{ccc}
\nu _w \mathcal{Y}_1 & V_Y \left(\mathcal{Y}_2+r_{Y}\mathcal{Y}_3\right) & V_X \left(\mathcal{Y}_4+r_{X}\mathcal{Y}_5\right) \\ 
 & V_Y\mathcal{Y}_6 & V_X \left(\mathcal{Y}_7+r\mathcal{Y}_5\right) \\
 & &V_X\mathcal{Y}_8
\end{array}
\right)
\label{lepton-mass}
\end{align}

where $\mathcal{Y}$ represents $3\times 3$ submatrices composed by Yukawa constants, while $r_{Y}=\nu _{w}/V_{Y}$, $r_{X}=\nu _{w}/V_{X}$ and $r=V_{Y}/V_{X}$ are VEV fractions. At first glance, we see in the diagonal blocks an evident hierarchical structure among a light $\underline{e}$ sector with masses of the order of the weak scale $\nu _w$, a heavy sector $\underline{E}$ at the 331 scale $V_Y$ and a superheavy sector $\underline{C}$ at the scale $V_X$. However, into each block there is not predictable hierarchical structures due to the unknown Yukawa matrices. The same structure in the mass matrices is found for the down-type quark sector and for the neutral leptons. However, in the up-type quark sector we obtain only a $3 \times 3$ matrix given by Eq. (\ref{Ap:up}), which exhibits two massless eigenstates, and only one quark get mass which can be identified with the top quark. In order to obtain hierarchical structures in each block of the lepton and down quark mass matrices, and to generate masses to the massless up-type quarks, we should explore one loop corrections.

\section{One-loop mass corrections}    

\begin{figure}[t]
\centering \includegraphics[scale=1]{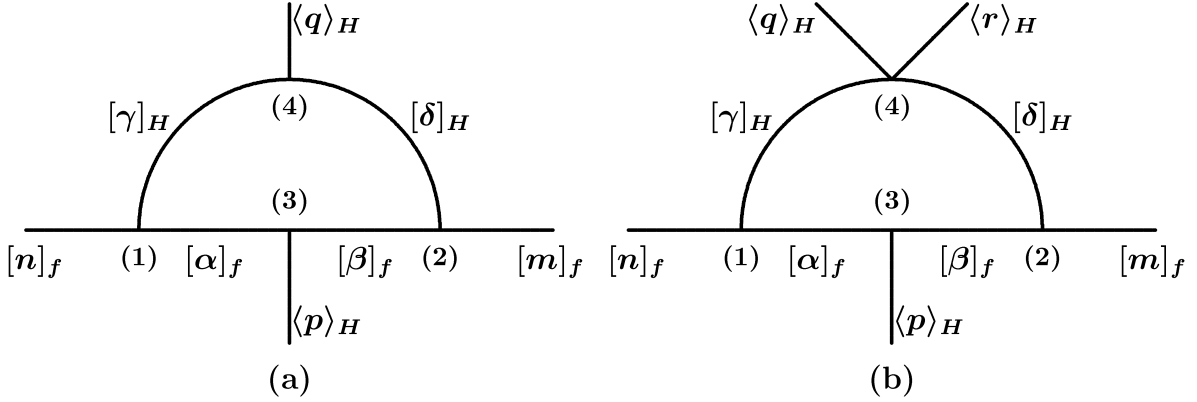}
\caption{One-loop diagrams for (a) two and (b) three Higgs representations evaluated in the VEVs, with $[n,m]_f$ external fermion tensors, $[\alpha ,\beta ]_f$ internal fermion tensors and $[\gamma , \delta ]_H$ internal Higgs fields}
\label{fig:figure1}
\end{figure}

In order to generate one-loop corrections, we use the following strategy. At tree level, we already generated fermion masses through renormalizable operators of the type shown in Eq. (\ref{yukawa}). However, we can induce higher-dimension operators through one-loop diagrams, which we classify in two types as shown schematically in figs. \ref{fig:figure1}. Each one-loop correction will generate mass operators that we approximate to the form:

\begin{align}
\nonumber &\langle p\rangle_H\langle q\rangle _H [n]_f[m]_f\rightarrow \frac{Kf\nu _p \nu_q}{16\pi ^2M^2}\psi _n \psi_m \sim \frac{K\nu _p \nu_q}{16\pi ^2M}\psi _n \psi_m\\ 
&\langle p\rangle_H\langle q\rangle _H\langle r\rangle_H [n]_f[m]_f \rightarrow \frac{K'\lambda \nu _p \nu_q \nu_ r}{16\pi ^2M^2}\psi _n \psi_m
\label{one-mass}
\end{align}  

where $\langle p\rangle _H$ label the VEV $\nu _p$ of a Higgs field of rank $[p]$. The constant $K$ ($K'$) contains products of Yukawa coupling constants from the vertices $(1),(2)$ and $(3)$ shown in the one-loop diagrams, and $M$ is associated to the mass of the heaviest internal particle which we assume is at the unification scale $M_{GUT}=V_I$. The constant $f$ is a characteristic trilinear coupling of the Higgs fields in the vertex $(4)$ of the first diagram, which has dimensions of mass and that we consider at order $f \sim M$ as shown in Eq. (\ref{one-mass}). Finally, the adimensional constant $\lambda $ is a representative quartic coupling for the vertex $(4)$ from the second diagram. In this work, we perform an analysis of the mass structure of the up-type quark sector. From the particle content in Eqs. (\ref{28-spectrum}) and (\ref{56-spectrum}), we choose those components that contains $u_{(n)aL}$ fields (embedded into $[n]_f=[2]$ and $[3]$ tensors), and construct all the one-loop diagrams of the form of Fig. \ref{fig:figure1}. Considering only dominant contributions, we obtain the structure $M_u=M_{u}^{0}+\delta m$, with $M_{u}^{0}$ the tree-level matrix from Eq. (\ref{Ap:up}) and $\delta m$ a correction matrix from the one-loop diagrams. Altogether, we obtain mass components of the form:

\begin{align}
\nonumber M_{u11} &\sim Y\nu _w \left[1+\frac{Y_{m}^{2}}{16\pi ^2}\frac{V_X}{M}\right], \\
\nonumber M_{u12}&=M_{u13}  \sim Y\nu _w \left\{\frac{y}{Y}+ \frac{(Y+y)Y_my_m\lambda }{16\pi ^2Y} \left[ \left(\frac{V_X}{M} \right)^2+\frac{V_XV_Y}{M^2} \right] \right\}, \\
\nonumber M_{u22}&=M_{u33} \sim Y\nu _w \left[ \frac{yy_mY_m\lambda }{16\pi ^2Y} \left(\frac{V_X}{M} \right)^2 \right], \\
 M_{u23} &\sim Y\nu _w \left\{ \frac{yy_mY_m\lambda }{16\pi ^2Y} \left[ \left(\frac{V_X}{M} \right)^2+\frac{V_XV_Y}{M^2} \right] \right\},
\label{full-up-mass}
\end{align} 

We can see that the most important contribution is the $11$ element, which contains the tree-level factor $1$ plus a correction of the order of $V_X/M \ll 1$. Then, a natural choice for this component is the top quark. The $12$ and $13$ components have the ratio $y/Y$, where $y$ is the $[2][3]\overline{[5]}$ coupling in Eq. (\ref{yukawa}) and $Y$ the $[2][2]\overline{[4]}$ coupling. In order to obtain the correct flavor structure in the mass matrix, we require that $y \ll Y$. Although this scheme in the Yukawa constants seems no natural, we could consider that the $[2][3]\overline{[5]}$ term comes from higher-dimensional operators, in a similar way that the one-loop diagrams above. Thus, it can indeed be assumed that the constant $y$ is not a tree-level coupling, but a small effective coupling of the order of $y \sim V_X/M$. As a first approximation, we can neglect small contributions in the elements from Eq. (\ref{full-up-mass}), and write down the following mass matrix for the up-type sector:

\begin{align}
M_u \approx Y\nu _w \left(
\begin{array}{ccc}
1 & \alpha & \alpha \\ 
 & \varepsilon ^2 & \varepsilon ^2+\varepsilon \delta \\
 & & \varepsilon ^2
\end{array}
\right),
\label{one-loop-mass}
\end{align}

with:

\begin{align}
\alpha = \frac{y}{Y} \sim \left(\frac{V_X}{M}\right), \hspace{1cm} \varepsilon &= \sqrt{\frac{yy_mY_m\lambda }{16\pi ^2Y}}\left(\frac{V_X}{M}\right), \hspace{1cm} \delta = \sqrt{\frac{yy_mY_m\lambda }{16\pi ^2Y}}\left(\frac{V_Y}{M}\right),
\end{align}

where the hierarchical relation $\delta \ll \varepsilon \lesssim \alpha \ll 1$ arise naturally from the symmetry-breaking scales in Eq. (\ref{scales}), with $M \sim V_I \gg V_X \gg V_Y$. Although it is not possible to give numerical predictions of the eigenvalues of the matrix in Eq. (\ref{one-loop-mass}) without a precise determination of the Yukawa coupling constants, a threefold hierarchy can be obtained. After diagonalizing, the matrix in Eq. (\ref{one-loop-mass}) displays nonzero eigenvalues which we can identify with the physical masses of the up-type quarks. Then, we obtain the following forms:

\begin{align}
|m_u| \sim Y\nu _w \varepsilon \delta, \hspace{1cm}|m_c| \sim 2Y\nu _w \alpha ^2 \hspace{1cm} |m_t| \sim Y \nu_ w ,
\end{align}   

from where we see that $|m_u| \ll |m_c| \ll |m_t|$. Using the experimental data $|m_u/m_c| \approx 1.9\times 10^{-3}$ and $|m_c/m_t| \approx 7.4\times 10^{-3}$, we obtain the constraints $\alpha \sim 0.06$ and $\varepsilon \delta \sim 1.4 \times 10^{-5}$. For example, if $\varepsilon \sim 0.01$, then $\delta \sim 1.4 \times 10^{-3}$, which respect the scheme $\delta \ll \varepsilon \lesssim \alpha \ll 1$ and predict the observed threefold hierarchy and ratios among up-type quark masses without any additional assumed condition on the Yukawa couplings. A similar analysis can be done in the down-type quark and lepton sector. A complete and more precise study of the mass structures and corrections will be done in future works.      

\section{Conclusions}

In this work we showed that it is possible to embed a 3-3-1 model into a single unified group SU(8) for three families, and simultaneously generate nontrivial flavor structures that leads to hierarchical schemes in the fermion mass matrices. Using a 3-3-1 "language" and appropriate branching rules, we obtained a decomposition of antisymmetric irreducible representations of SU(8), where light SM-type, heavy 3-3-1-type and superheavy-type fermions can be identified. Introducing an appropriate Higgs sector and from the symmetry-breaking mechanism, we generated tree-level masses to the three sets of fermions at different energy scales. We showed that hierarchical mass structures are obtained if we include one-loop corrections. For the up-type sector, we obtained the ratios $|m_u/m_c| \approx 1.9\times 10^{-3}$ and $|m_c/m_t| \approx 7.4\times 10^{-3}$ with few assumptions on free parameters and without any hierarchical requirements on the Yukawa couplings. However, in order to study in more detail the consequences and predictive power of this model, we should extend the above analysis, for example, studying the renormalization group equations, how to generate hierarchical schemes in the lepton ($l$) and down-type quark ($d$) sector that can account for the ratios $m_l/m_d$, etc., which we intend to do in future works.              

\vspace{0.2cm}
This work was supported by Colciencias,

\section*{Appendix}

\appendix

\section{Electric charges\label{apendixA}}

From Eqs. (\ref{U1-gener})-(\ref{elec-charge}) and using the decomposition in (\ref{irrep-bran}), we obtain the following electric charges:

\begin{align}
\nonumber \psi _{(m)A} \rightarrow Q_{[a]}&=-1/3, \\ \nonumber
 Q_{[4]}&=-\alpha _1 - \alpha _2, \\ \nonumber 
 Q_{[i]}&=\left(-1,0,-1/3 +\alpha _1 + \alpha _3 /3 \right), \\ 
 Q_{[8]}&= -\alpha _1 - \alpha _3 
\label{Ap:charges1}
\end{align}

\begin{align}
\nonumber \psi ^{[AB]} \rightarrow Q^{[ab]}&=-2\left(\alpha _1 - \alpha _2 /3 \right), \\ \nonumber
 Q^{[a4]}&=-2\left(\alpha _1 + \alpha _2 /3 \right), \\ \nonumber
 Q^{[ai]}&=\left(2/3+ \alpha _2 /3 + \alpha _3 /3,-1/3+ \alpha _2 /3 + \alpha _3 /3,-1/3+ \alpha _2 /3 + \alpha _3 /3  \right), \\ \nonumber
 Q^{[a8]}&= \alpha _2 /3 - \alpha _3 \\ \nonumber
 Q^{[4i]}&=\left(2/3- \alpha _2 + \alpha _3 /3, -1/3- \alpha _2 + \alpha _3 /3, -1/3- \alpha _2 + \alpha _3 /3 \right), \\ \nonumber
 Q^{[48]}&=-\alpha _2 - \alpha _3 , \\ \nonumber
 Q^{[ij]}&=\left( 1/3+2 \alpha _1 + 2\alpha _3 /3, 1/3+2\alpha _1 + 2\alpha _3 /3, -2/3+2\alpha _1 + 2\alpha _3 /3 \right) \\ 
 Q^{[i8]}&=\left(2/3+ 2\alpha _1 - 2\alpha _3 /3, -1/3+ 2\alpha _1 - 2\alpha _3 /3, -1/3+ 2\alpha _1 - 2\alpha _3 /3 \right) 
\label{Ap:charges2}
\end{align}

\begin{align}
\nonumber \psi ^{[ABC]} \rightarrow Q^{[abc]}&=-3\left( \alpha _1 - \alpha _2 /3 \right) \\ \nonumber
 Q^{[ab4]}&=-3\alpha _1 - \alpha _2 /3, \\ \nonumber
 Q^{[abi]}&=\left(2/3- \alpha _1 + 2\alpha _2 /3 + \alpha _3 /3, -1/3- \alpha _1 + 2\alpha _2 /3 + \alpha _3 /3, -1/3- \alpha _1 + 2\alpha _2 /3 + \alpha _3 /3 \right), \\ \nonumber
 Q^{[ab8]}&=-\alpha _1 + 2\alpha _2 /3 - \alpha _3, \\ \nonumber
 Q^{[a4i]}&=\left(2/3-\alpha _1 - 2\alpha _2 /3 + \alpha _3 /3, -1/3- \alpha _1 - 2\alpha _2 /3 + \alpha _3 /3, -1/3- \alpha _1 - 2\alpha _2 /3 + \alpha _3 /3 \right), \\ \nonumber
 Q^{[a48]}&=- \alpha _1 - 2\alpha _2 /3 - \alpha _3 , \\ \nonumber
 Q^{[aij]}&= \left(1/3+ \alpha _1 + \alpha _2 /3 + 2\alpha _3 /3, 1/3+ \alpha _1 + \alpha _2 /3 + 2\alpha _3 /3, -2/3+ \alpha _1 + \alpha _2 /3 + 2\alpha _3 /3 \right), \\ \nonumber
 Q^{[ai8]}&=\left(2/3+ \alpha _1 + \alpha _2 /3 - 2\alpha _3 /3, -1/3+ \alpha _1 + \alpha _2 /3 - 2\alpha _3 /3, -1/3+ \alpha _1 + \alpha _2 /3 -2 \alpha _3 /3 \right), \\ \nonumber
 Q^{[4ij]}&=\left(1/3+ \alpha _1 - \alpha _2 + 2\alpha _3 /3, 1/3+ \alpha _1 - \alpha _2 + 2\alpha _3 /3, -2/3+ \alpha _1 - \alpha _2 + 2\alpha _3 /3 \right), \\ \nonumber
 Q^{[4i8]}&=\left(2/3+ \alpha _1 - \alpha _2 - 2\alpha _3 /3, -1/3+ \alpha _1 - \alpha _2 - 2\alpha _3 /3, -1/3+ \alpha _1 - \alpha _2 - 2\alpha _3 /3 \right), \\ \nonumber
 Q^{[ijk]}&=3\alpha _1 + \alpha _3 , \\ 
 Q^{[ij8]}&=\left(1/3+ 3\alpha _1 - \alpha _3 /3, 1/3+ 3\alpha _1 - \alpha _3 /3, -2/3+ 3\alpha _1 - \alpha _3 /3 \right) 
\label{Ap:charges3}
\end{align}

where $A=1,..,8$ are SU(8) indices, $a,b,c=1,..3$ $SU(3)_c$ indices, $i,j,k=5,..7$ are $SU(3)_L$ indices, and 8,9 are singlets. The charges distributed in triplets run over $i-$type indices.

\section{Fermion masses\label{appendixAA}}

\subsection{Charged lepton masses}

From the particle assignation in Eqs. (\ref{8-spectrum})-(\ref{56-spectrum}) and the VEVs in (\ref{VEV}), we obtain in the Yukawa structures in Eq. (\ref{yukawa}) the fermion mass matrices. According to the scale scheme adopted by Ec. (\ref{scales}), we can reduce the VEVs variables in ec. (\ref{VEV}) choosing the following scales: $\nu _{w}=\nu _{jw}$ for all $j=1,..9$, $V_{Y}=V_{jY}$ for all $j=1,..4$, and $V_c=V_X=V_{jX}$ for $j=1,2$. For the charged lepton sector, we obtain the following mass matrix at tree level:

\begin{align}
 M^{0}_{l}&=\left(
\begin{array}{ccc}
\left(\mathcal{M}^{0}_{l}\right)_{SM} & \left(\mathcal{M}^{0}_{l}\right)_{31/SM} & \left(\mathcal{M}^{0}_{l}\right)_{SH/SM} \\ 
 & \left(\mathcal{M}^{0}_{l}\right)_{31} & \left(\mathcal{M}^{0}_{l}\right)_{SH/31} \\
 & &\left(\mathcal{M}^{0}_{l}\right)_{SH}
\end{array}
\right),
\label{Ap:char-lep}
\end{align}
 
where the indices $SM$, $31$ and $SH$ label $3 \times 3$ blocks corresponding to light Standard Model ($SM$) lepton basis, the new 331 ($31$) heavy lepton basis and superheavy ($SH$) lepton basis, respectively, while $31/SM$, $SH/SM$ etc. label mixing blocks. The $SM$ block form the following symmetric matrix in the light basis $\underline{e}=(e^{\pm}_{(1)},e^{\pm}_{(2)},e^{\pm}_{(3)})$:

\begin{align}
\left(\mathcal{M}^{0}_{l}\right)_{SM}=\nu _w \left(
\begin{array}{ccc}
Y_1 & Y_2-y_1 & Y_3+y_1 \\ 
 & -y_2 & y_2-y_3 \\
 & & y_3
\end{array}
\right).
\label{Ap:light-char}
\end{align} 

For the 331 charged leptons $\underline{E}=\left(E^{\pm}_{(1)},E^{\pm}_{(2)},E^{\pm}_{(3)}\right)$ we obtain:

\begin{align}
\left(\mathcal{M}^{0}_{l}\right)_{31}=V_Y \left(
\begin{array}{ccc}
Y_4 & Y_5-y_4 & Y_6+y_4 \\ 
 & -y_5 & y_5-y_6 \\
 & & y_6
\end{array}
\right).
\label{Ap:31-char}
\end{align}

For the superheavy charged leptons $\underline{C}=\left(C^{\pm}_{(1)},C^{\pm}_{(2)},C^{\pm}_{(3)}\right)$: 

\begin{align}
\left(\mathcal{M}^{0}_{l}\right)_{SH}=V_X \left(
\begin{array}{ccc}
-Y_7 & Y_7-Y_8 & -Y_9-y_7 \\ 
 & Y_8 & Y_9-y_8 \\
 & & -y_9
\end{array}
\right)
\label{Ap:SH-char}
\end{align}

We can see in the above expressions an strong hierarchical scheme impose by the VEV of the Higgs, where the mass blocks follow the relations $SM \ll 31 \ll SH$.    For the mixing blocks, we obtain:

\begin{align}
\left(\mathcal{M}^{0}_{l}\right)_{31/SM}=V_Y \left(
\begin{array}{ccc}
Y_1+r_{Y}Y_{4} & Y_2-r_{Y}y_{4} & Y_3+r_{Y}y_{4} \\ 
-y_1+r_{Y}Y_{5} & -y_2-r_{Y}y_{5} & -y_3+r_{Y}y_{5} \\
y_1+r_{Y}Y_{6} & y_2-r_{Y}y_{6} & y_3+r_{Y}y_{6}
\end{array}
\right)
\label{Ap:3M-char}
\end{align}

\begin{align}
\left(\mathcal{M}^{0}_{l}\right)_{SH/SM}=V_X \left(
\begin{array}{ccc}
-Y_1+r_{X}Y_7 & -Y_2-r_{X}y_7 & -Y_3+r_{X}y_7 \\ 
Y_1+r_{X}Y_8 & Y_2-r_{X}y_8 & Y_3+r_{X}y_8 \\
-y_1+r_{X}Y_9 & -y_2-r_{X}y_9 & -y_3+r_{X}y_9
\end{array}
\right)
\label{Ap:HM-char}
\end{align}

\begin{align}
\left(\mathcal{M}^{0}_{l}\right)_{SH/31}=V_X \left(
\begin{array}{ccc}
-Y_4+rY_7 & -Y_5-ry_7 & -Y_6+ry_7 \\ 
Y_4+rY_8 & Y_5-ry_8 & Y_6+ry_8 \\
-y_4+rY_9 & -y_5-ry_9 & -y_6+ry_9
\end{array}
\right)
\label{Ap:H3-char}
\end{align}

with $r_{Y}=\nu _{w}/V_{Y}$, $r_{X}=\nu _{w}/V_{X}$ and $r=V_{Y}/V_{X}$.

\subsection{Down-type quark masses}

Down-type quark mass matrix follows from a similar analysis, obtaining at tree-level:

\begin{align}
 M^{0}_{d}&=\left(
\begin{array}{ccc}
\left(\mathcal{M}^{0}_{d}\right)_{SM} & \left(\mathcal{M}^{0}_{d}\right)_{31/SM} & \left(\mathcal{M}^{0}_{d}\right)_{SH/SM} \\ 
 & \left(\mathcal{M}^{0}_{d}\right)_{31} & \left(\mathcal{M}^{0}_{d}\right)_{SH/31} \\
 & &\left(\mathcal{M}^{0}_{d}\right)_{SH}
\end{array}
\right)
\label{Ap:down}
\end{align}
 
where in the basis $\underline{d}=\left(d_{(1)},d_{(2)},d_{(3)}\right)$, $\underline{D}=\left(D_{(1)},D_{(2)},D_{(3)}\right)$ and $\underline{J}=\left(J_{(1)},J_{(2)},J_{(3)}\right)$, we obtain respectively:  

\begin{align}
\nonumber \left(\mathcal{M}^{0}_{d}\right)_{SM}=\nu _w &\left(
\begin{array}{ccc}
Y_1 & Y_2+y_1 & Y_3+y_1 \\ 
 & y_2 & y_2+y_3 \\
 & & y_3
\end{array}
\right),
\left(\mathcal{M}^{0}_{d}\right)_{31}=V_Y \left(
\begin{array}{ccc}
Y_4 & Y_5+y_4 & Y_6+y_4 \\ 
 & y_5 & y_5+y_6 \\
 & & y_6
\end{array}
\right) \\ 
&\left(\mathcal{M}^{0}_{d}\right)_{SH}=V_X \left(
\begin{array}{ccc}
Y_7 & Y_7+Y_8 & Y_9+y_7 \\ 
 & Y_8 & Y_9+y_8 \\
 & & y_9
\end{array}
\right),
\label{Ap:light-down}
\end{align} 

for the diagonal blocks, while for the mixing blocks:

\begin{align}
\left(\mathcal{M}^{0}_{d}\right)_{31/SM}=V_Y \left(
\begin{array}{ccc}
Y_1+r_{Y}Y_4 & Y_2+r_{Y}y_4 & Y_3+r_{Y}y_4 \\ 
y_1+r_{Y}Y_5 & y_2+r_{Y}y_5 & y_3+r_{Y}y_5 \\
y_1+r_{Y}Y_6 & y_2+r_{Y}y_6 & y_3+r_{Y}\left(y_4+y_6\right)
\end{array}
\right),
\label{Ap:3M-down}
\end{align}

\begin{align}
\left(\mathcal{M}^{0}_{d}\right)_{SH/SM}=V_X \left(
\begin{array}{ccc}
Y_1+r_{X}Y_7 & Y_2+r_{X}y_7 & Y_3+r_{X}y_7 \\ 
Y_1+r_{X}Y_8 & Y_2+r_{X}y_8 & Y_3+r_{X}y_8 \\
y_1+r_{X}Y_9 & y_2+r_{X}y_9 & y_3+r_{X}y_9
\end{array}
\right),
\label{Ap:HM-down}
\end{align}

\begin{align}
\left(\mathcal{M}^{0}_{d}\right)_{SH/31}=V_X \left(
\begin{array}{ccc}
Y_4+rY_7 & Y_5+ry_7 & Y_6+ry_7 \\ 
Y_4+rY_8 & Y_5+ry_8 & Y_6+ry_8 \\
y_4+rY_9 & y_5+ry_9 & y_6+ry_9
\end{array}
\right),
\label{Ap:H3-down}
\end{align}

where, again $r_{Y}=\nu _{w}/V_{Y}$, $r_{X}=\nu _{w}/V_{X}$ and $r=V_{Y}/V_{X}$.

\subsection{Up-type quark masses}

For the up-type quark sector, we obtain only a light matrix in the basis $\underline{u}=(u_{(1)},u_{(2)},u_{(3)})$:

\begin{align}
M^{0}_{u}=\nu _w \left(
\begin{array}{ccc}
Y & y & y \\ 
 & 0 & 0 \\
 & & 0
\end{array}
\right).
\label{Ap:up}
\end{align}

The diagonalization of this matrix leads to two massless quarks, which is not observed in the nature. In order to find the missing components to obtain non-vanishing masses, we should explore one-loop diagrams such as explained in Sec. 5.  

\subsection{Neutral lepton masses}

The spectrum obtained in Eqs. (\ref{8-spectrum})-(\ref{56-spectrum}) contains in total 46 neutral majorana leptons, which represents a huge number of fields to deal with in a unified way. Thus, instead of trying to write down a $46 \times 46$ mass matrix let us separate the neutral basis according how they transform under the SM sub-group $(SU(3)_c,SU(2)_L)$:

\vspace{0.3cm}

1.-) \textit{Singlets (1,1):} In the spectrum in Eq. (\ref{8-spectrum}) we identify the following neutral lepton singlets: 

\begin{align}
\nonumber \psi _{(m)4}&=\left(X^{0}_{(1)},X^{0}_{(2)},X^{0}_{(3)};R^{0}_{(1)},R^{0}_{(2)},R^{0}_{(3)};B^{0}_{(1)},B^{0}_{(2)},B^{0}_{(3)}\right)_L \\
\nonumber \psi _{(m)7}&=\left(N^{0}_{(1)},N^{0}_{(2)},N^{0}_{(3)};F^{0}_{(1)},F^{0}_{(2)},F^{0}_{(3)};G^{0}_{(1)},G^{0}_{(2)},G^{0}_{(3)}\right)_L \\
 \psi _{(m)8}&=\left(Y^{0}_{(1)},Y^{0}_{(2)},Y^{0}_{(3)};P^{0}_{(1)},P^{0}_{(2)},P^{0}_{(3)};I^{0}_{(1)},I^{0}_{(2)},I^{0}_{(3)}\right)_L.
\label{singlet_1}
\end{align} 

In (\ref{28-spectrum}) we see:

\begin{align}
\psi ^{[47]}=M^{0}_{(1)L}; \hspace{1cm} \psi ^{[48]}=O^{0}_L; \hspace{1cm} \psi ^{[78]}=M^{0}_{(2)L}.
\label{singlet_2}
\end{align}

Finally in (\ref{56-spectrum}):

\begin{align}
\psi ^{[478]}=M^{0}_{(3)L}
\label{singlet_3}
\end{align}
\vspace{0.3cm}

2.-) \textit{Doublets (1,2):} There are other neutral leptons which are embedded into weak doublets together with a charged lepton, for example the $(e^{-}_{(n)},\nu _{(n)})$ doublet. Although we separate the neutral part from the charged lepton, we will denote these fields as doublet-type neutral leptons. So in the spectrum in Eq. (\ref{8-spectrum}) we identify the following doublet-type neutral leptons: 

\begin{align}
 \psi _{(m)6}&=\left(\nu _{(1)},\nu _{(2)},\nu _{(3)};E^{0}_{(1)},E^{0}_{(2)},E^{0}_{(3)};C^{0}_{(1)},C^{0}_{(2)},C^{0}_{(3)}\right)_L,
\label{doublet_1}
\end{align} 

in (\ref{28-spectrum}):

\begin{align}
\psi ^{[46]}=L^{0}_{(1)L}; \hspace{1cm} \psi ^{[68]}=L^{0}_{(2)L}; \hspace{1cm} \psi ^{[67]}=\tilde{E}^{0}_{(1)L},
\label{doublet_2}
\end{align}

and in (\ref{56-spectrum})

\begin{align}
\psi ^{[468]}=L^{0}_{(3)L}; \hspace{1cm} \psi ^{[467]}=\tilde{E}^{0}_{(2)L}; \hspace{1cm} \psi ^{[678]}=\tilde{E}^{0}_{(3)L}.
\label{doublet_3}
\end{align}

It is direct to verify that there are 46 fields in Eqs. (\ref{singlet_1})-(\ref{doublet_3}). We construct all the mass matrix from the Yukawa terms in (\ref{yukawa}), which in terms of the above basis are written as singlet-singlet (\textit{SS}), doublet-doublet (\textit{DD}) and singlet-doublet (\textit{SD}) interactions:

\begin{align}
\nonumber -i\mathcal{L}^{SS}&=\overline{\left(\psi _{(m)7}\right)^c}M_1\left[\psi _{(l)4}-\psi _{(l)8}\right]+\overline{\left(\psi _{(m)7}\right)^c}M_2\left[\psi ^{[47]}-\psi ^{[78]}\right]+\overline{\left(\psi _{(m)4}\right)^c}M_3\psi _{(l)8} \\
\nonumber &+\overline{\left(\psi _{(m)4}\right)^c}M_4\psi ^{[47]}+\overline{\left(\psi _{(m)4}\right)^c}M_5 \psi ^{[478]}-\overline{\left(\psi _{(m)4}\right)^c}M_2\psi ^{[48]}-\overline{\left(\psi _{(m)8}\right)^c}M_4\psi ^{[78]} \\
 &+\overline{\left(\psi _{(m)8}\right)^c}M_5\psi ^{[478]}+\overline{\left(\psi _{(m)8}\right)^c}M_2\psi ^{[48]}+h.c,
\label{SS-mass}
\end{align}

\begin{align}
\nonumber -i\mathcal{L}^{DD}&=\overline{\left(\psi _{(m)6}\right)^c}M_2\left[\psi ^{[46]}-\psi ^{[68]}\right]+\overline{\left(\psi _{(m)6}\right)^c}M_6\psi ^{[468]}+\overline{\left(\psi _{(m)6}\right)^c}M_4\psi ^{[67]}+\overline{\left(\psi _{(m)6}\right)^c}M_5\psi ^{[678]} \\ &+\overline{\left(\psi _{(m)6}\right)^c}M_7 \psi _{(l)6}+h.c,
\label{DD-mass}
\end{align}

\begin{align}
\nonumber -i\mathcal{L}^{DS}&=\overline{\left(\psi _{(m)6}\right)^c}M_8\psi _{(l)7}+\overline{\left(\psi _{(m)6}\right)^c}M_9\psi _{(l)4}+\overline{\left(\psi _{(m)6}\right)^c}M_{10}\psi _{(l)8}+\overline{\left(\psi _{(m)6}\right)^c}M_6\psi ^{[478]} \\
\nonumber &+\overline{\left(\psi _{(m)7}\right)^c}M_{11}\psi ^{[67]}+\overline{\left(\psi _{(m)7}\right)^c}M_{12}\psi ^{[678]}-\overline{\left(\psi _{(m)4}\right)^c}M_{11}\psi ^{[46]}-\overline{\left(\psi _{(m)4}\right)^c}M_{12}\psi ^{[468]} \\
\nonumber &+\overline{\left(\psi _{(m)8}\right)^c}M_2\psi ^{[68]}-\overline{\left(\psi _{(m)8}\right)^c}M_{12}\left[\psi ^{[468]}+\psi ^{[678]}\right] +\overline{\left(\psi ^{[47]}\right)^c}M_{13}\psi ^{[68]}-\overline{\left(\psi ^{[78]}\right)^c}M_{13}\psi ^{[46]} \\ &-\overline{\left(\psi ^{[48]}\right)^c}M_{13}\psi ^{[67]}+h.c,
\label{DS-mass}
\end{align}

where $M_{K}$ represents $9 \times 9$ and $9 \times 1$ matrices. In a short form, the components of each mass matrix have the form (for $m,l=1,...,9$):

\begin{align}
\nonumber &\left(M_1\right)_{ml}=-a_{ml}V_Y;& \hspace{0.5cm} &\left(M_2\right)_{m}=-Y_{m}V_X;& \hspace{0.5cm} &\left(M_3\right)_{ml}=a_{ml}V_X;& \\
\nonumber &\left(M_4\right)_{m}=Y_{m}V_Y;& \hspace{0.5cm} &\left(M_5\right)_{m}=y_{m}V_Y;& \hspace{0.5cm} &\left(M_6\right)_{m}=-y_{m}V_X;& \\
\nonumber &\left(M_7\right)_{ml}=-\left(\delta _{m5}y_{l}+y_{m}\delta _{l5}\right)V_Y;& \hspace{0.5cm} &\left(M_8\right)_{ml}=\left(a_{ml}+\delta _{m5}y_{l}\right)\nu _w;& \hspace{0.5cm} &\left(M_9\right)_{ml}=\left(-a_{ml}+\delta _{m5}y_{l}\right)\nu _w;& \\
\nonumber &\left(M_{10}\right)_{ml}=a_{ml}\nu _w;& \hspace{0.5cm}  &\left(M_{11}\right)_{m}=-Y_{m}\nu _w;& \hspace{0.5cm} &\left(M_{12}\right)_{m}=-y_{m}\nu _w;& \\
&\left(M_{13}\right)=-Y\nu _w.&
\label{neutral-mass} 
\end{align}

\end{document}